\documentclass[aps,prl,twocolumn,showpacs,superscriptaddress,10pt]{revtex4-1}
\usepackage{bm,amsmath,amssymb,latexsym,mathrsfs,enumerate}
\usepackage[mathcal]{euscript}
\usepackage[normalem]{ulem}
\usepackage{xcolor}

\usepackage{graphicx, subfigure}

\usepackage[disable]{todonotes}
\newcommand{\delKH}[1]{\todo[inline]{#1}}

\begin{document}

\title{Transformation of spin current by antiferromagnetic insulators}

\author{Roman Khymyn}

\email{khiminr@gmail.com}

\affiliation{Department of Physics, Oakland University, Rochester,
Michigan 48309, USA}

\author{Ivan Lisenkov}

\affiliation{Department of Physics, Oakland University, Rochester,
Michigan 48309, USA}

\affiliation{Institute of Radio-engineering and Electronics of RAS,
Moscow 125009, Russia}

\author{Vasil S. Tiberkevich}
\affiliation{Department of Physics, Oakland University, Rochester,
Michigan 48309, USA}

\author{Andrei N. Slavin}
\affiliation{Department of Physics, Oakland University, Rochester,
Michigan 48309, USA}

\author{Boris A. Ivanov}
\affiliation{Institute of Magnetism, NASU and MESYSU, Kiev 03142,
Ukraine}

\begin{abstract}
It is demonstrated theoretically that a thin layer of an anisotropic antiferromagnetic (AFM) insulator can effectively conduct spin current through the excitation of a pair of evanescent AFM spin wave modes. The spin current flowing through the AFM is not conserved due to the interaction between the excited AFM modes and the AFM lattice, and, depending on the excitation conditions, can be either attenuated or enhanced. When the phase difference between the excited evanescent modes is close to $\pi/2$, there is an optimum AFM thickness for which the output spin current reaches a maximum, that can significantly exceed the magnitude of the input spin current. The spin current transfer through the AFM depends on the ambient temperature and increases substantially when temperature approaches the Neel temperature of the AFM layer.
\end{abstract}

\maketitle

Progress in modern spintronics critically depends on finding novel media that can  serve as effective conduits of spin angular momentum over large distances with minimum losses \cite{Spintronics,Spintronics1, Demidov}. The mechanism of spin transfer is reasonably well-understood in ferromagnetic (FM) metals \cite{spincurrentGeneral, metals} and insulators \cite{Demidov, spincurrentGeneral, ferromagnets, ferromagnets1, ferromagnets2, Demokritov}, but  there are only very few theoretical papers describing spin current in  antiferromagnets (AFM) (see, e.g., \cite{Tserkovnyak2}).

\begin{figure}[t!]
\begin{center}
        \includegraphics*[width=0.9\linewidth]{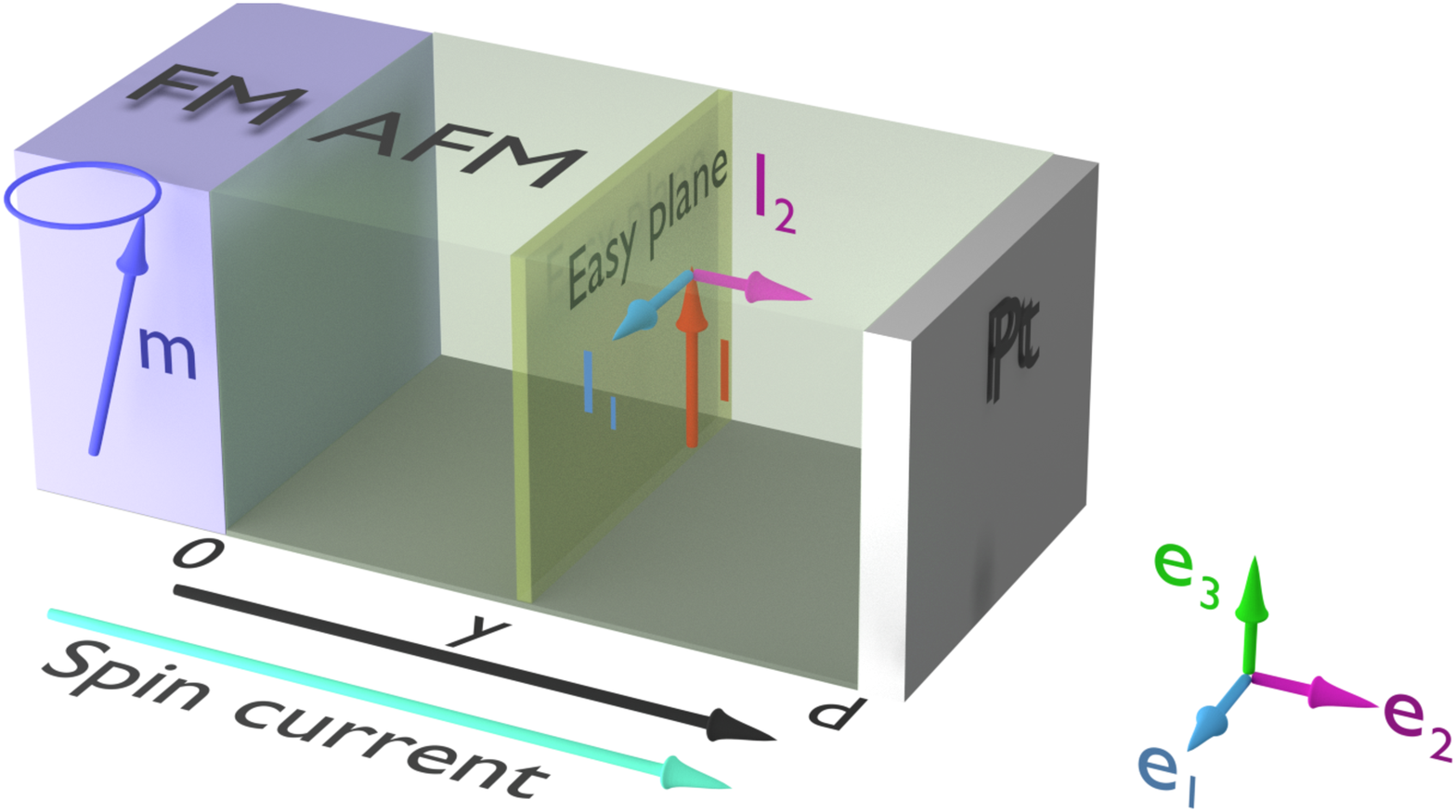}
\end{center}
        \caption{Sketch of the model of spin current transfer through an AFM insulator based on the experiment \cite{AFMExperiment}.
        The FM layer excites spin wave excitations in the AFM layer. The output spin current (at the  AFM/Pt interface) is detected by the Pt layer through the inverse spin Hall effect (ISHE).}
\label{Model}
\end{figure}

The recent experiments \cite{AFMExperiment, AFMExperiment1, Saitoh}
have demonstrated that a thin layer of a dielectric AFM (NiO, CoO)
could  effectively conduct spin current.  The transfer of spin
current was studied in the FM/AFM/Pt trilayer structure (see
Fig.~\ref{Model}). The FM layer driven in ferromagnetic resonance (FMR) excited spin current
in a thin layer of AFM, which was
detected in the adjacent Pt film using the inverse spin Hall effect
(ISHE). It was also found in \cite{Saitoh} that the spin current
through the AFM depends on the ambient temperature and goes through
a maximum near the Neel temperature $T_N$. The most intriguing
feature of the experiments was the  fact that for
a certain optimum thickness of the AFM layer ($\sim5$~nm) the
detected spin current had a maximum \cite{AFMExperiment,
AFMExperiment1},
which could be even higher than in the absence of the AFM spacer \cite{AFMExperiment1}.
The spin current transfer
in the reversed geometry, when the spin current flows from the  Pt
layer  driven by DC current through the AFM spacer into a
microwave-driven FM material has been reported recently in
\cite{Tserkovnyak1}.

The experiments \cite{AFMExperiment, AFMExperiment1,Tserkovnyak1, Saitoh} posed a fundamental question of the mechanism of the apparently rather effective spin current transfer through an AFM dielectric. A possible mechanism of the spin transfer through an \emph{easy-axis} AFM has been recently proposed in \cite{Tserkovnyak2}.
However, this uniaxial model can not explain the non-monotonous dependence of the transmitted spin current on the AFM layer thickness and the apparent ``amplification'' of the spin current
seen in the experiments \cite{AFMExperiment, AFMExperiment1}
performed with the \emph{bi-axial} NiO AFM layer \cite{AFMNuclScat}.

In this Letter, we propose a possible mechanism of spin current
transfer through anisotropic AFM dielectrics, which may explain all
the
peculiarities of the experiments \cite{AFMExperiment, AFMExperiment1,Tserkovnyak1}.
Namely, we show that the spin current
can be effectively carried by the driven {\em evanescent}
spin wave excitations, having frequencies that are much lower than
the frequency of the AFM resonance. We demonstrate that the angular
momentum
exchange
between the spin subsystem and the AFM lattice plays a crucial role
in this process,
and may lead to the {\em enhancement} of the spin
current
inside the AFM layer.

We consider a model of a simple AFM having two magnetic sublattices
with the partial saturation magnetization $M_s$.
%
The
distribution of the magnetizations of each sublattice can be
described by the vectors $\mathbf{M}_1$ and $\mathbf{M}_2$, $|\mathbf{M}_1|=|\mathbf{M}_1|=M_s$. We use a
conventional approach for describing the AFM dynamics by
introducing the vectors of antiferromagnetism ($\mathbf{l}$) and magnetism ($\mathbf{m}$) \cite{sigmamodel,
sigmamodelAfleck, Kosevich, IvanovSatoh}:
\begin{equation}
    \mathbf{l}=(\mathbf{M}_1-\mathbf{M}_2)/(2 M_s),\quad
    \mathbf{m}=(\mathbf{M}_1+\mathbf{M}_2)/(2 M_s).
\end{equation}
%
Assuming that all the magnetic fields
are smaller then
the exchange field $H_{ex}$ and neglecting the bias magnetic field, that is used to saturate the FM
layer, the effective AFM Lagrangian can be written as \cite{sigmamodel, Kosevich, IvanovSatoh}:
\begin{multline}
    \mathcal{L}=\mu \left[\left(\partial \mathbf{l}/\partial
t\right)^2 -c^2 \left(\partial \mathbf{l}/\partial
y\right)^2 \right]-
W_a-W_s\delta(y).
\label{Lagrangian}
\end{multline}
Here $\mu=M_s/(\gamma^2 H_{ex})$,
$\gamma$ is the gyromagnetic ratio,
$c$ is the speed of the AFM spin waves ($c\simeq33$~km/s in NiO),
and $W_a = M_s \mathbf{l} \cdot (\hat{\mathbf{H}}^a \cdot \mathbf{l})$ is the energy of the anisotropy,
defined by the matrix of the anisotropy fields
$\hat{\mathbf{H}}^a= \text{diag}(H^a_{1},H^a_{2},0)$ with the
diagonal $(j, j)$ components $H^a_{j}=2 M_s\beta_j$ ($\beta_j$
is the anisotropy constant along the $j$-th axis). The equilibrium
direction of the AFM vector $\mathbf{l}_0 =
\mathbf{e}_3$ lies along the $\mathbf{e}_3$ axis.


The exchange coupling between the FM and AFM layers is modeled in Eq.~(\ref{Lagrangian}) by the surface energy term
$W_s=E_s\left[(\mathbf{m}_{\text{FM}} \mathbf{\cdot m})+\alpha(\mathbf{m_{\text{FM}}} \cdot \mathbf{l})\right]$, where $E_s$ is the
surface energy density,
$\mathbf{m}_\text{FM}$ is the unit vector of FM layer magnetization. We assumed that the net magnetization of the AFM at the FM/AFM interface layer could be partially non-compensated, and this ``non-compensation'' is characterized by a dimensionless parameter $\alpha$ ($0 < \alpha <1$).

\begin{figure}[t!]
 \begin{center}
  \includegraphics*[width=0.9\linewidth]{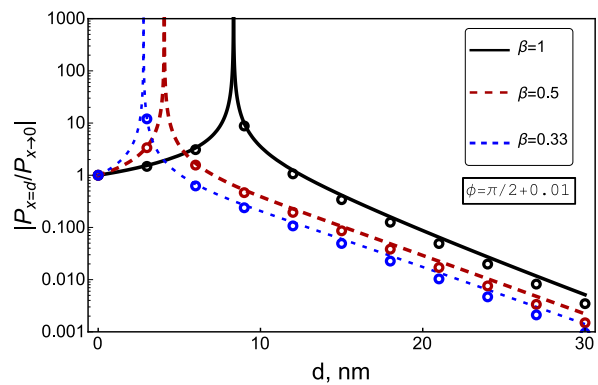}
  \end{center}
   \caption{Spin current transfer factor of the AFM layer as a function of the AFM layer thickness for different values of the spin mixing conductance parameter $\beta$. The lines show the case of zero damping ($\Delta\omega=0$), while the circles correspond to the AFMR linewidth of
	$\Delta \omega/2\pi=69$~GHz \cite{Sievers}.}
 \label{PdP0}
\end{figure}

The dynamical equation for the AFM vector $\textbf{l}$ follows from the Lagrangian Eq.~(\ref{Lagrangian}) and can be written as
\begin{equation}
\partial^2 \mathbf{l}/\partial t^2 +\Delta \omega \,\partial \mathbf{l}/\partial t -c^2 \, \partial^2
\mathbf{l}/\partial y^2 + \hat{\mathbf{\Omega}} \cdot \mathbf{l}=
\mathbf{f}(t)\delta(y), \label{oscequation}
\end{equation}
where $\Delta \omega$ is the phenomenological damping parameter equal to the AFM resonance linewidth ($\Delta\omega/2\pi\approx~69$GHz for NiO \cite{Sievers}). Note, that the damping-related decay length $\lambda_G = 2c/\Delta\omega \approx 150$~nm is much larger than the typical AFM thickness. Therefore, below we shall neglect damping except in Fig.~\ref{PdP0}, where the comparison of AFM spin currents in conservative and damped cases is presented.
The matrix
$\hat{\mathbf{\Omega}}=\text{diag}(\omega_1^2,\omega_2^2,0),$ and
$\omega_j = \gamma\sqrt{H_{ex}H^a_{j}}$, $j=1,2$, are the frequencies
of the AFM resonance. In the case of NiO
the two AFM resonance frequencies are substantially
different: $\omega_1/2\pi \simeq 240$~GHz and $\omega_2/2\pi \simeq
1.1$~THz \cite{AFMNuclScat}. We shall show below that the
difference between the AFM resonance frequencies is crucially important for the spin current transfer
through the AFM.

The driving force in Eq.~(\ref{oscequation}) $\mathbf{f}(t)=-(\delta W_s /\delta\mathbf{l})/(2\mu)$, localized at the FM/AFM interface, describes AFM excitation by the precessing FM magnetization. In the absence of this term Eq.~(\ref{oscequation}) describes two branches of the
eigen-excitations of the AFM with dispersion relations
$\omega_j(\mathbf{k}) = \sqrt{\omega_j^2 + c^2k^2}$. These
propagating AFM spin waves
have minimum frequencies
$\omega_j$ which are much higher than the excitation frequency ($9.65$~GHz in
Ref.~\cite{AFMExperiment}) and, therefore, can not be
responsible for the spin current transfer.

The presence of the
FM layer, however,
qualitatively changes the situation, as the driving
force $\mathbf{f}(t)$ excites \emph{evanescent}
AFM spin wave modes at the frequency of the FM layer resonance (FMR), that is well below any of
the AFMR frequencies $\omega_j$.
The profiles of the evanescent AFM modes
can be easily found from  Eq.~(\ref{oscequation}):
\begin{equation}
\mathbf{l}_j(t,y) =\mathbf{e}_j\left[\mathcal{A}_j e^{-y/\lambda_j}
+\mathcal{B}_j e^{y/\lambda_j} \right]e^{- i \omega t} +
\mathrm{c.c.},\quad j=1,2, \label{solution}
\end{equation}
where  $\omega$ is the excitation frequency,
\begin{equation}
\lambda_j = c\big/\sqrt{\omega_j^2-\omega^2} \label{lambda}
\end{equation}
is the penetration depth for the $j$-th evanescent mode, and complex coefficients
$\mathcal{A}_{j}$, $\mathcal{B}_{j}$
are determined by the boundary conditions at the FM/AFM and AFM/Pt
interfaces.
%
The interfacial driving force $\mathbf{f}(t)\delta(y)$ excites the AFM
vector $\mathbf{l}(t, y=0)$ at the FM/AFM interface:
\begin{equation}
    \mathbf{l}(t, y=0) = \mathbf{e}_3 + \left[
            \left(a_1\mathbf{e}_1 + a_2\mathbf{e}_2\right)e^{-i\omega t} + \mathrm{c.c.}
        \right].
    \label{l0}
\end{equation}
The complex amplitudes $a_1$ and $a_2$ depend on the vector
structure of the magnetization precession in the FM layer (see Supplementary Materials for details), which
opens a way to experimentally control the input spin current in the AFM,
and to directly verify our theoretical predictions.  Thus, if the FM
layer is magnetized along one of the AFM anisotropy axes
$\mathbf{e}_{1,2}$, the microwave magnetization component along that
axis will be zero and the corresponding complex amplitude $a_{1,2}$
in Eq.~(\ref{l0}) will vanish. On the other hand, if the FM layer is
magnetized along the AFM equilibrium axis $\mathbf{e}_3$, both
amplitudes $a_1$ and $a_2$ will be non-zero with the phase shift
$\phi = \mathrm{arg}(a_1/a_2) \approx \pi/2$ between them.

At the AFM/Pt interface ($y=d$) we adopt a simple form of the
boundary conditions that were used previously for the description of
spin current at the AFM/Pt \cite{Bratas} and FM/Pt \cite{Tserkovnyak} interfaces:
\begin{equation}
    P (y=d) = \beta\, c\, L(y=d),\label{bc-Pt}
\end{equation}
where $P$ is the current of the ${\mathbf{e}_3}$-component
of the spin angular momentum and $L$ is the corresponding angular momentum density inside the AFM:
\begin{equation}
    P=2 \mu c^2 \mathbf{e}_3 \cdot [{\partial
\mathbf{l}/\partial y} \times\mathbf{l}],\quad
    L =-2 M_s \gamma^{-1} \, \mathbf{e}_3\cdot \mathbf{m},
 \label{flux}
\end{equation}
and $\beta$ is a dimensionless constant having magnitude in the
range from 0 to 1 and being physically determined by the spin mixing
conductance at the AFM/Pt interface \cite{Tserkovnyak}. The case
$\beta=0$ corresponds to the conservative situation of a complete absence of the angular momentum flux, while the case $\beta=1$ describes a
``transparent'' boundary, when the angular momentum freely moves
across the AFM/Pt boundary without any reflection.

Using Eqs.~(\ref{flux}), the boundary conditions Eq.~(\ref{bc-Pt})
can be rewritten as explicit conditions on the vector of
antiferromagnetism $\mathbf{l}$ as $\beta\, \partial
\mathbf{l}/\partial t= - c\, \partial \mathbf{l}/\partial y$. This
equation and Eq.~(\ref{l0}) allow one to find all four
coefficients $\mathcal{A}_j$, $\mathcal{B}_j$ in
Eq.~(\ref{solution}), and one can find the explicit expression for
the spin current $P(y)$ inside the AFM layer:
\begin{equation}
    P(y) = 4\mu c^2|a_1a_2|\mathrm{Re}\left[Q(y)e^{-i\phi}\right],
    \label{result}
\end{equation}
where
\begin{multline}
    Q(y) = \frac{(e^{-y/\lambda_1}+q_1e^{y/\lambda_1})(e^{-y/\lambda_2}-q_2^*e^{y/\lambda_2})}
            {(1+q_1)(1+q_2^*)\lambda_2}\\
    -\frac{(e^{-y/\lambda_1}-q_1e^{y/\lambda_1})(e^{-y/\lambda_2}+q_2^*e^{y/\lambda_2})}
            {(1+q_1)(1+q_2^*)\lambda_1}\,.
\end{multline}
Here $q_j = e^{2i\psi_j-2d/\lambda_j}$ and $\psi_j =
\mathrm{arctan}(\beta\omega\lambda_j/c) \approx
\beta\omega/\omega_j$. Eq.~(\ref{result}) is the central result of
this paper that allows one to find the spin current carried by the
evanescent spin wave modes in an AFM layer. \delKH{Below we shall analyze
the main features of the spin current transfer through an AFM
dielectric that are described by Eq.~(\ref{result}).}

Using Eq.~(\ref{result}), we estimated the ISHE voltage \cite{SaitohVoltage} for an FM/AFM/Pt structure (NiFe/ NiO/Pt structure)  with the AFM layer  having thickness $d$  that is  smaller than the penetration depths  $\lambda_j$ of both evanescent AFM modes for the following parameters: $\beta=1$, $E_s=3.3$~erg/cm$^2$~\cite{Celinski}, FM precession angle $\sin \theta=0.01$, Pt spin Hall angle $\theta_{\mathrm{SH}}=0.05$~\cite{AFMExperiment1}, Pt length $5$~mm~\cite{AFMExperiment}, Pt thickness $10$~nm. We obtained $V_{\text{ISHE}}=40$~mV for the fully uncompensated AFM ($\alpha=1$) and $V_{\text{ISHE}}=5$~nV for the fully compensated AFM ($\alpha=0$). The first value is close to the ISHE voltage measured in Ref.~\cite{AFMExperiment}. A partial interfacial magnetization of  the antiferromagnetic NiO in this case is confirmed by the XRD scan performed in \cite{AFMExperiment}. The calculated ISHE voltage for the compensated AFM is closer to the experimental value obtained in Ref.~\cite{AFMExperiment1}.

Now we shall analyze the main features of the spin current transfer through an AFM dielectric that are described by Eq.~(\ref{result}). First, one can see that the spin current $P$ is proportional to the product $|a_1a_2|$ of the amplitudes of both excited evanescent spin wave modes, and this current is completely absent if only one of the modes is excited. This is explained by the fact that each of the modes Eq.~(\ref{solution}) is linearly polarized, and, therefore, can not alone carry any angular momentum.

\begin{figure}
 \begin{center}
  \includegraphics*[width=0.9\linewidth]{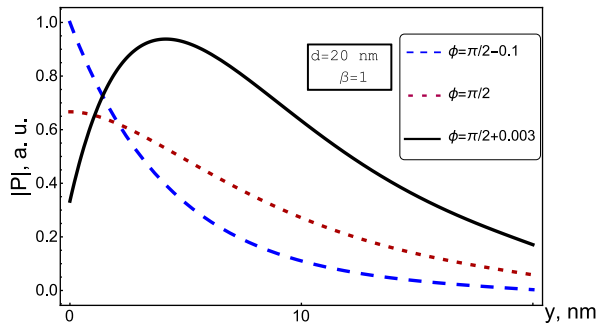}
  \end{center}
   \caption{Spatial distribution of the spin current $P(y)$ inside the AFM layer for different phase shifts $\phi$ between the two evanescent AFM spin wave modes calculated from Eq.~(\ref{result}).}
 \label{Pvsx}
\end{figure}

Second, the spin current in the AFM layer depends on the position
$y$ inside the AFM layer, i.e., it is \emph{not conserved}. This is
a direct consequence of the assumed bi-axial anisotropy of the AFM
material, which allows for the transfer of the angular momentum
between the spin sub-system and the crystal lattice of the AFM
layer.
This effect is a magnetic analogue of the optical effect of birefringence \cite{biref},
where the spin angular momentum of light is dynamically changed during its propagation in a birefringent medium.



In the case of a uniaxial anisotropy \cite{Tserkovnyak2} ($\lambda_1
= \lambda_2 = \lambda$) Eq.~(\ref{result}) can be simplified to
\begin{equation}
    P = \frac{16\mu c^2}{\lambda}\frac{\mathrm{Im}(q)}{|1+q|^2}|a_1 a_2|\sin\phi,
    \label{isotropic}
\end{equation}
and the spin current \emph{is conserved} across the whole AFM layer.

Another peculiarity of Eq.~(\ref{result}) and Eq.~(\ref{isotropic})
is that the spin current $P$ depends on the phase shift $\phi$
between the two excited evanescent AFM spin wave modes $\mathbf{l}_1$ and
$\mathbf{l}_2$: $P \propto \cos(\phi-\Phi(y))$, where $\Phi(y) =
\mathrm{arg}(Q(y))$. The maximum spin current {\em at a given
position $y$} inside the AFM layer is achieved at $\phi = \Phi(y)$.
Since the AFM phase shift $\Phi(y)$, in general, depends on the
position $y$ inside the AFM layer, for any particular thickness $d$
of the AFM layer it is possible to choose the excitation phase shift
$\phi$ that would maximize the output spin current $P(y=d)$, while
the input spin current $P(y\rightarrow 0)$ could be quite low. In
such a case the additional angular momentum is taken from the
crystal lattice of the AFM. This shows that, in principle, the AFM
dielectrics can serve as ``amplifiers'' of a spin current.

Fig.~\ref{Pvsx} shows the spatial profiles of the spin current
density in a relatively thick AFM layer (thickness $d=20$~nm). This
dependence is drastically different for different phase shifts
$\phi$ between the excited evanescent spin wave modes. While for
$\phi < \pi/2$ the spin current exponentially and monotonically
decays inside the  AFM layer (dashed blue line in Fig.~\ref{Pvsx}),
for $\phi > \pi/2$ (solid black line in Fig.~\ref{Pvsx}) it initially
increases at relatively small $y$ due to the angular momentum flow from
the AFM crystal lattice to its spin subsystem. At larger values of
$y$, the spin current decays exponentially due to the
decay of the excited evanescent spin wave modes.

\begin{figure}
 \begin{center}
  \includegraphics*[width=0.9\linewidth]{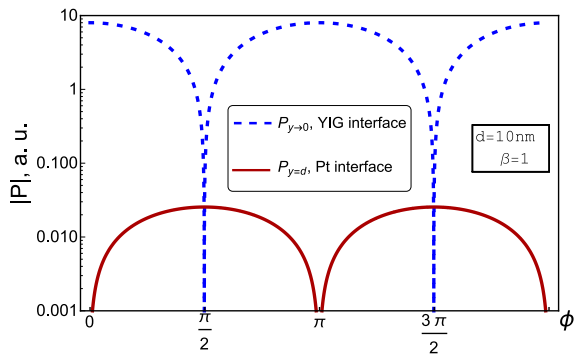}
  \end{center}
   \caption{Dependence of the input (dashed blue line) and output (solid red line) spin currents through the AFM layer on the phase shift $\phi$ between the modes.}
 \label{PvsFi}
\end{figure}

Fig.~\ref{PvsFi} demonstrates the dependences of the spin
current on the phase shift $\phi$  at both interfaces
FM/AFM (input spin current) and AFM/Pt (output spin current).
It is clear from Fig.~\ref{PvsFi} that the output spin current is shifted by $\sim\pi/2$ relative to the input
spin current, and, for the phase shift $\phi \approx \pi/2$, the output
spin current could have a maximum magnitude when the input spin current is almost completely absent.
This means, that at such a value
of the phase shift between the evanescent spin wave modes practically all the
output spin current is generated as a result of interaction between
the  magnetic subsystem of the AFM layer and its crystal lattice.
Thus, the AFM layer acts as a \emph{source} of the spin current. On
the other hand, at the phase shift of $\phi \approx 0$ or $\phi \approx \pi$, the
situation is opposite, as the input spin current is
practically lost inside the AFM, and the AFM layer acts as a spin current \emph{sink}.

Thus, we showed, that a thin layer of AFM, driven by a constant flow of microwave energy from the FM layer, is able to transform the angular momentum of a crystal lattice into the spin current and vice versa. The
described transfer of the angular momentum from the lattice to the
spin system has a simple analog not only as a birefringence in optics, but also
in mechanics: a mechanical oscillator which consists of a mass suspended on two perpendicular
springs with different stiffness attached  to a fixed rectangular
frame. The displacement of the mass from its equilibrium position in
the frame center along the direction of one of the orthogonal
springs results in the linearly polarized oscillations along this
direction, without any transfer of the angular momentum from the
frame to the oscillating mass. In contrast, the linear displacement
of the mass in a \emph{diagonal} direction results in the
\emph{rotation }of the mass around its equilibrium position, and the
angular momentum necessary for this rotation is taken from the frame
(see animations in Suppl. Mater.).

The ratio of the output spin current to the input one (the spin
current transfer factor) is shown in Fig.~\ref{PdP0} for different
values of the constant $\beta$, i.e., for the different values of the
spin mixing conductance at the AFM/Pt interface. This dependence has
a sharp maximum at the thickness of a few nanometers, where the
input current is rather low, and the AFM layer acts as a source of a spin
current. With the further increase of the AFM layer thickness
the transfer ratio is exponentially decreasing, while the position of the maximum shifts
to the right with the increase of the spin mixing conductance at the AFM/Pt interface.
As one can see from Fig.~\ref{PdP0}, the presence of the damping has little influence on the spin current,
because the spatial decay of the amplitudes due to the evanescent character of the modes $\mathbf{l}_{1,2}$ is dominant.

The above presented results were obtained for the parameters of
bulk NiO at low temperature. However, it is well known that such
important parameters of AFM substances as the anisotropy
constants and Neel temperature in thin AFM films could be
substantially smaller than in bulk crystals (see, e.g., \cite{Alders}).
Thus, the penetration depths of the evanescent spin wave modes
Eq.~(\ref{lambda}), determined at a given driving frequency $\omega$  by
the AFM anisotropy constants, would significantly depend on the
thickness and the temperature of the AFM layer. Particularly, with
the increase of temperature the AFMR frequencies
$\omega_j$ would decrease and approach zero at the Neel temperature
\cite{Sievers}. In accordance with Eq.~(\ref{lambda}), this means that
the penetration depth of the evanescent spin wave modes will
increase  substantially when the temperature approaches the Neel temperature of the  AFM layer.
This increase of the spin current transferred through the  AFM layer is clearly seen in
the experiments \cite{Saitoh}.

In conclusion, we demonstrated  that the spin current can be
effectively transmitted through thin dielectric AFM layers by a pair of externally excited evanescent AFM spin wave modes.
In the case of AFM materials with bi-axial anisotropy the transfer of angular momentum between
the spin subsystem and the crystal lattice  of the AFM can lead to the enhancement or decrease of the transmitted spin current, depending on the phase relation between the excited evanescent spin wave modes. Our results explain all the qualitative features of the recent experiments
\cite{AFMExperiment, AFMExperiment1,Tserkovnyak1, Saitoh}, in particular, the existence of an optimum thickness
of the AFM layer, for which the output current could reach a maximum
value which is higher than the spin current magnitude in the absence of the
AFM spacer, and the increase of the transmitted spin current at the temperatures close to the Neel temperature of the AFM layer.

This work was supported in part by the Grant ECCS-1305586 from the National Science Foundation of the USA, by the contract from the US Army TARDEC, RDECOM, by DARPA MTO/MESO grant N66001-11-1-4114 and by the Center for NanoFerroic Devices (CNFD) and the Nanoelectronics Research Initiative (NRI).

\end{document}